\documentclass[12pt]{article}
\usepackage{cite}
\usepackage{amsmath, amsthm, amssymb, bbm, dsfont,relsize, diagrams}
\usepackage{ifpdf, slashed}
\ifpdf
  \usepackage[pdftex]{graphicx}
  \usepackage{epstopdf}
\else
  \usepackage[dvips]{graphicx}
\fi
\parskip=\baselineskip
\def\Bbb{\mathbb}
\def\Tr{{\rm Tr}}

\font\teneurm=eurm10 \font\seveneurm=eurm7 \font\eighteurm=eurm8 \font\fiveeurm=eurm5
\newfam\eurmfam
\textfont\eurmfam=\teneurm \scriptfont\eurmfam=\seveneurm
\scriptscriptfont\eurmfam=\fiveeurm

\font\teneusm=eusm10 \font\seveneusm=eusm7 \font\fiveeusm=eusm5
\newfam\eusmfam
\textfont\eusmfam=\teneusm \scriptfont\eusmfam=\seveneusm
\scriptscriptfont\eusmfam=\fiveeusm

\font\tencmmib=cmmib10 \skewchar\tencmmib='177
\font\sevencmmib=cmmib7 \skewchar\sevencmmib='177
\font\fivecmmib=cmmib5 \skewchar\fivecmmib='177
\newfam\cmmibfam
\textfont\cmmibfam=\tencmmib \scriptfont\cmmibfam=\sevencmmib
\scriptscriptfont\cmmibfam=\fivecmmib

\def\g{\text{{\teneurm g}}}

\def\sg{\text{{\eighteurm g}}}

\def\5d{{\mathrm{5d}}}
\def\4d{{\mathrm{4d}}}
\def\16{{\bf 16}}
\def\1{{\bf 1}}
\def\2{{\bf 2}}
\def\3{{\bf 3}}
\def\4{{\bf 4}}

\def\d{{\mathrm d}}

\def\bar{\overline}
\def\tilde{\widetilde}

\def\R{{\Bbb{R}}}\def\Z{{\Bbb{Z}}}

\def\N{{\mathcal N}}
\def\hat{\widehat}
\font\teneurm=eurm10 \font\seveneurm=eurm7 \font\fiveeurm=eurm5
\newfam\eurmfam
\textfont\eurmfam=\teneurm \scriptfont\eurmfam=\seveneurm
\scriptscriptfont\eurmfam=\fiveeurm

\font\teneusm=eusm10 \font\seveneusm=eusm7 \font\fiveeusm=eusm5
\newfam\eusmfam
\textfont\eusmfam=\teneusm \scriptfont\eusmfam=\seveneusm
\scriptscriptfont\eusmfam=\fiveeusm

\font\tencmmib=cmmib10 \skewchar\tencmmib='177
\font\sevencmmib=cmmib7 \skewchar\sevencmmib='177
\font\fivecmmib=cmmib5 \skewchar\fivecmmib='177
\newfam\cmmibfam
\textfont\cmmibfam=\tencmmib \scriptfont\cmmibfam=\sevencmmib
\scriptscriptfont\cmmibfam=\fivecmmib

\numberwithin{equation}{section}

\def\d{\mathrm d}
\def\bar{\overline}

\def\Z{{\Bbb Z}}

\def\X{{\mathcal X}}
\def\CC{{\mathcal C}}
\def\F{{\mathcal F}}

\def\W{{\mathcal W}}

\def\bar{\overline}

\def \TT{{\sf T}}
\begin{document}
\begin{titlepage}
\vskip 1.5in
\begin{center}
{\bf\Large{More on Gopakumar-Vafa formula:\vskip0.1cm coefficients $\F_0$ and $\F_1$}}
\vskip0.5cm 
{Mykola Dedushenko} 
\vskip.5cm
 {\small{\textit{Joseph Henry Laboratories, Princeton University, Princeton NJ USA 08540}}}
\end{center}
\vskip.5cm
\baselineskip 16pt
\begin{abstract}
In Type IIA compactified on a Calabi-Yau threefold,  the genus zero and one terms of the Gopakumar-Vafa (GV) formula describe F-terms that are related to genus zero and one topological amplitudes.  While for higher-genus terms $\F_\sg,\g\ge 2$, the contribution of a light hypermultiplet can be computed via a sum over Kaluza-Klein harmonics,  as has
been shown in a recent paper,  for $\g\leq 1$, 
the sum diverges and it is better to compute $\F_0$ and $\F_1$ directly in five-dimensional field theory. Such a computation is presented here.
\end{abstract}
\date{December, 2014}
\end{titlepage}
\section{Introduction}
In Type IIA superstring theory compactified on a Calabi-Yau manifold $Y$, a series of F-terms in the  $d=4, N=2$ effective supergravity action are known to be exactly computable. Originally discovered from the topological string side \cite{BC} and identified as certain physical superstring amplitudes \cite{AGNT}, they were later reinterpreted by Gopakumar and Vafa \cite{GV1, GV2} using the space-time effective theory and lifting to M-theory. The latter approach was recently reexamined in \cite{GVrev}.

In terms of $N=2$ superspace, the interactions that are computed by the GV formula are $-i\int\d^4x\d^4\theta \F_\sg(\X)(\W^2)^\sg$, where $\W_{\mu\nu}$ and $\X$ are Weyl and vector superfields. The superfields used here naturally appear in the formulation of $d=4, N=2$ supergravity, in which one first constructs superconformal gravity and then breaks the extra part of its gauge supergroup (dilatations, special conformal transformations, special supersymmetries and $SU(2)\times U(1)$ R-symmetry) by explicitly choosing a certain gauge slice. The relevant concepts will be briefly reviewed in section \ref{sugra}.

The Gopakumar-Vafa formula gives an expression for $\F_\sg$ coefficients in terms of the spectrum of BPS states in M-theory compactified on $Y\times S^1$, where $S^1$ is the M-theory circle. The space-time derivation of this expression is based on computing the contribution to the Wilsonian effective action due to  5d particles winding the M-theory circle. Moreover, only trajectories with non-zero winding number 
have to be considered.  Trajectories with zero winding number naively give an ultraviolet-divergent contribution, but as 
 is explained in \cite{GVrev}, this contribution should be regarded as part of the 5d effective action and need not be calculated.  Only
  a few terms in the 5d effective action are actually relevant to the GV formula, and these terms are known because of their
  relation to anomalies.

As emphasized in \cite{GVrev}, BPS states that are massive in five dimensions are more naturally treated as particles in deriving
their contribution to the GV formula, while those that are massless (or anomalously light) in five dimensions are more naturally treated
as fields.  Particle-based and field theory computations were performed in \cite{GVrev}, but the field theory computation left a gap,
which we will treat here.

The field theory computation in \cite{GVrev} was based on turning on a constant graviphoton background, as suggested in
\cite{GV1,GV2}, summing over Kaluza-Klein harmonics, and reducing to Schwinger's computation of the effective action of a charged
particle in a 4d magnetic field. (It is also necessary, technically, to perturb slightly away from a flat metric on $\R^4$.) This method works nicely for $\F_\sg$ with $\g\geq 2$, but for $\g\leq 1$, there are two problems.  One problem is that the sum over Kaluza-Klein
harmonics that is supposed to determine $\F_1$ is divergent, and from a 4d point of view it is not clear how to regularize it properly.
One would expect a similar divergence in the Kaluza-Klein sum for $\F_0$, but actually there is an additional problem for $\F_0$:
it does not contribute to the effective action in the background considered in \cite{GVrev}, so to determine it one would need
to perform a one-loop computation in a less convenient background.

On the other hand, to compute $\F_0$ and $\F_1$, there is no need to turn on a graviphoton background.  $\F_0$ contributes
to the kinetic energy of the scalar fields in vector multiplets, and $\F_1$ contributes a Weyl squared interaction.  So $\F_0$ and $\F_1$
can be computed by calculating one-loop contributions to the scalar and graviton two-point functions.   There is then no need to turn
on a graviphoton background, and that being so, there is also no advantage to expanding in KK harmonics.  The purpose of the
KK expansion had been that this simplifies the one-loop computation in the presence of a graviphoton field.  

Instead, the two-point functions can be naturally
computed directly in five dimensions.  The advantage of this is that there is no problem with ultraviolet divergences: any ultraviolet
divergence would be a five-dimensional integral of some local, gauge-invariant operator, and as explained in \cite{GVrev}, terms
of this form do not contribute to $\F_\sg$.  Even better, the computation of the two-point functions can be expressed as a sum over
5d trajectories of various winding numbers, and the appropriate elimination of UV divergences is accomplished by just throwing away
the contribution of winding number 0. 

In section \ref{sugra}, we briefly review relevant facts about supergravity. In section \ref{Fg}, we discuss some properties of the F-terms we are computing and constrain the expected form of the answers for $\F_0$ and $\F_1$ based on symmetries. In section \ref{F0}, we describe the 1-loop computation of $\F_0$, emphasizing a subtle point in the relation between the deformation of the Kahler metric for the vector multiplet scalars and the deformation of $\F_0$. In section \ref{F1}, we describe the 1-loop computation of $\F_1$. Then some related discussions are added in section \ref{disc}.

\section{5d and 4d supergravities with 8 supercharges}\label{sugra}
\subsection{5d supergravity coupled to $U(1)$ vector multiplets}\label{sug5d}
The full and detailed construction of 5d supergravity can be found in \cite{5dsugra}. Here we outline the main features that will be relevant for us. We use the notations and conventions of \cite{GVrev}.

The $U(1)$ vector multiplet in 5d has a gauge field, a real scalar and a spinor. The physical scalars are described by $n$ constrained scalars $h^I,\,I=1\dots n$ satisfying the constraint:
\begin{equation}
\label{hyps}
\CC_{IJK}h^Ih^Jh^K=1,
\end{equation}
where $\CC_{IJK}$ is a symmetric constant real tensor. In the case of Calabi-Yau compactifications of M-theory, these $h^I$ parametrize the Kahler cone of the Calabi-Yau $Y$, and the tensor $\CC_{IJK}=\frac{1}{6}\int_Y\omega_I\wedge\omega_J\wedge\omega_K$ contains intersection numbers (here $\omega_I$ are a basis in a degree-$(1,1)$ cohomology). One also introduces $h_I=\CC_{IJK}h^Jh^K$ and $a_{IJ}=-3\CC_{IJK}h^K + \frac{9}{2}h_Ih_J$. In the Calabi-Yau case, $a_{IJ}=\frac{1}{4}\int_Y \omega_I\wedge *\omega_J$ is a natural metric on the Kahler cone. When pulled back on the hypersurface defined by (\ref{hyps}), this metric defines the kinetic energy of physical scalars. In addition to $n$ constrained scalars $h^I$, there are also $n$ gauge fields $V^I$, $I=1\dots n$. One specific linear combination of the field strengths, namely $\TT=\sum_{I=1}^n h_I\d V^I$, is a graviphoton of the supergravity multiplet. So in total, adding corresponding fermions, we have a supergravity multiplet and $n-1$ vector multiplets.

Of course, in the case of the Calabi-Yau compactification of 11d supergravity, there are also hypermultiplets present (see \cite{Mth5}), but including them does not change much.

The bosonic part of the action can be written as:
\begin{equation}
\label{act5d}
S =\int \Big[{1\over 2}R^{(5)} + {3\over 2}\CC_{IJK}h^I \partial_M h^J \partial^M h^K -{1\over 4}a_{IJ} (dV^I\cdot dV^J)\Big]\textrm{Vol}-{1\over 2}\CC_{IJK}V^I\wedge dV^J \wedge dV^K,
\end{equation}
where ${\rm Vol}$ is a volume form on spacetime.
\subsection{4d supergravity coupled to $U(1)$ vector multiplets}\label{sug4d}
The formulation of 4d Poincare supergravity that we rely on is based on superconformal gravity, which is gauge-equivalent to Poincare supergravity in the sense that partial gauge fixing of the superconformal theory gives Poincare supergravity. This naturally comes with an $N=2$ superspace. Chirtal superfields of weight $2$ under dilations can be considered as possible F-terms in the superspace action of conformal supergravity. Given some superspace interaction, say an F-term $\int\d^4x\d^4\theta\, \Phi$, to find the corresponding terms in the Poincare supergravity action, one has to not only integrate over Grassmann coordinates $\theta$, but also impose all gauge-fixing constraints that reduce the superconformal gauge group to the super Poincare.

Two superconformal matter multiplets, the compensators, disappear in this gauge-fixing. One usually chooses a vector multiplet and a hypermultiplet for this role (see \cite{n2sug} for details). Thus to build an $N=2$ Poincare supergravity coupled to $n$ vector multiplets, one starts with $N=2$ superconformal gravity coupled to $n+1$ vector multiplets and 1 hypermultiplet.

An $N=2$ vector multiplet in 4d contains a complex scalar, a vector and a doublet of spinors. Such multiplets are described by reduced chiral superfields $\X^\Lambda$, $\Lambda=0\dots n$ (see \cite{supf, structn2, extconf}), whose lowest components $X^\Lambda$ are complex scalars, while the highest components are $-\frac{1}{6}(\varepsilon_{ij}\bar\theta^i\sigma^{\mu\nu}\theta^j)^2 D_\mu D^\mu \bar X^\Lambda$ and involve derivatives of complex conjugate scalars (because of the non-holomorphic constraint satisfied by reduced chiral superfields). Coupling of vector multiplets is described by the holomorphic prepotential $\F_0(\X)$ (see \cite{n2sug}), which has to be homogeneous of degree 2 to define a term in the Lagrangian of conformal supergravity:
\begin{equation}
\label{F_act}
-i\int\d^4x\d^4\theta\, \F_0(\X) + c.c.
\end{equation}
One introduces the usual notations $F_\Lambda = \partial \F_0/\partial X^\Lambda$, $F_{\Lambda\Sigma}=\partial^2\F_0/(\partial X^\Lambda\partial X^\Sigma)$ etc. Another useful notation is:
\begin{equation}
N_{\Lambda\Sigma}=2{\rm Im}\, F_{\Lambda\Sigma}.
\end{equation}
Superspace expression (\ref{F_act}) implies the kinetic term for conformal scalars:
\begin{equation}
\int\d^4x \sqrt{g} N_{\Lambda\Sigma}D_\mu X^\Lambda D^\mu\bar{X}^\Sigma,
\end{equation}
where the derivatives are covariant with respect to the superconformal gauge group. In order to get the kinetic energy of scalars of Poincare supergravity, one has to use a gauge condition which fixes dilatational symmetry of conformal supergravity. This usually has a form of some constraint on the superconformal scalars $X^\Lambda$. The freedom to perform local dilatations in conformal supergravity corresponds to the freedom to Weyl-rescale metric in Poincare supergravity. The standard gauge choice \cite{n2sug}, which guarantees that the Poincare theory emerges written in the Einstein frame, is
\begin{equation}
\label{standard}
N_{\Lambda\Sigma}X^\Lambda \bar{X}^\Sigma=-1.
\end{equation}
It is usually supplemented by the $U(1)$ R-symmetry gauge, which we pick as $iX^0>0$. A convenient choice of independent holomorphic scalars is:
\begin{equation}
Z^I=\frac{X^I}{X^0},~~I=1\dots n.
\end{equation}
The standard gauge choice (\ref{standard}) implies the following expression for $|X^0|^2$ in terms of other fields:
\begin{equation}
|X^0|^2 = \frac{1}{Y},~~Y=-N_{\Lambda\Sigma}Z^\Lambda\bar{Z}^\Sigma.
\end{equation}
In this case, the kinetic energy takes the form:
\begin{equation}
\label{kinstand}
Y^{-1}\mathfrak{M}_{IJ}\partial_\mu Z^I\partial^\mu \bar{Z}^J,\quad
\mathfrak{M}_{IJ} = N_{IJ} - (N_{I\Lambda}\bar{X}^\Lambda)(N_{J\Sigma}X^\Sigma),
\end{equation}
and $Y^{-1}\mathfrak{M}_{IJ}$ is actually a Kahler metric:
\begin{equation}
Y^{-1}\mathfrak{M}_{IJ}=\frac{\partial}{\partial Z^I}\frac{\partial}{\partial\bar{Z}^J}\ln{Y}.
\end{equation}

For completeness, we also write expression for the kinetic term of the gauge fields. It is independent of the dilatational gauge and is given by:
\begin{equation}
-\frac{i}{4}\N_{\Lambda\Sigma}F^{\Lambda +}_{\mu\nu}F^{\Sigma+\mu\nu}+c.c.,
\end{equation}
where $F^{\Lambda+}_{\mu\nu}$ are the self-dual parts of the field strengths of the elementary gauge fields $A^\Lambda_\mu$, and $\N$ is a scalar-dependent matrix:
\begin{equation}
\label{Ncal}
\N_{\Lambda\Sigma}=\bar{F}_{\Lambda\Sigma} + i\frac{(NX)_\Lambda (NX)_\Sigma}{(X,NX)}.
\end{equation}

\subsection{Reduction from 5d to 4d}\label{red5to4}
Kaluza-Klein reduction of the 5d supergravity of section \ref{sug5d} gives $N=2$ supergravity in 4d, which can be described in terms of the fields of section \ref{sug4d}. As was shown in \cite{GVrev}, only $\F_0$ and $\F_1$ interactions can arise in this way. However, $\F_1$ requires inclusion of higher-derivative terms in the 5d action. If we start from the action with no more than 2 derivatives in 5d, we will get only the classical prepotential in 4d:
\begin{equation}
\label{class}
\F_0^{\rm cl} = -\frac{1}{2}\frac{\CC_{IJK}X^IX^JX^K}{X^0}.
\end{equation}

If the index $\mu$ represents 4d coordinates and the fifth coordinate (along the circle) is $y$, the 5d metric in the Kaluza-Klein reduction takes the form:
\begin{equation}
\d s^2 = e^{-\sigma}g_{\mu\nu}\d x^\mu \d x^\nu + e^{2\sigma}(dy + B_\mu \d x^\mu)^2.
\end{equation}

Let the 5d vectors have non-zero holonomies in the circle direction $V^I_y=\alpha^I$. The following field redefinitions relate 5d vectors $V^I_M,~I=1\dots n$ and 5d scalars $h^I$ to the 4d vectors $A^\Lambda,~\Lambda=0\dots n$ and 4d scalars $Z^I$:
\begin{align}
A^I_\mu &= V^I_\mu - \alpha^I B_\mu,\cr
A^0_\mu &= -B_\mu,\cr
Z^I &= \alpha^I + i e^\sigma h^I.
\end{align}
We do not discuss reduction in the fermionic sector, as the formulas are more complicated and are not needed in this paper. More details can be found in Appendix A of \cite{GVrev}. Another reference on dimensional reduction of this particular supergravity is \cite{jord}.

By the classical dimensional reduction of 5d supergravity, as described above, we get $d=4, N=2$ supergravity in the standard gauge $N_{\Lambda\Sigma}X^\Lambda \bar{X}^\Sigma=-1$, so that
\begin{equation}
Y=4 e^{3\sigma}.
\end{equation}

\section{Some properties of $\F_\sg$}\label{Fg}
\subsection{Shift symmetries}\label{shift}
As was explained in \cite{GVrev}, the 4d effective action should be invariant under the shift symmetries $\alpha^I \to \alpha^I + n^I$, where $n^I\in \Z,~I=1\dots n$. This is evident because the only physical effect of holonomies $\alpha^I$ is through the factor $e^{2\pi i q_I\alpha^I}$ which is acquired by the particle of charge $\{q_I\}$ winding once around the circle. Indeed, if one considers a certain amplitude in a 4d theory, this amplitude is represented (in a particle description) as a sum over trajectories of particles, some of which can wind the extra circular dimension and thus can acquire such a factor. Thus all physical answers in 4d depend only on $e^{2\pi i \alpha^I}$ and should be invariant under $\alpha^I \to \alpha^I + n^I$. This is equivalent to the symmetry:
\begin{equation}
Z^I \to Z^I + n^I.
\end{equation}
 Now we note that the only way the 5d BPS miultiplet action will depend on $\alpha^I$ and $h^I$ is through the linear combinations $q_I\alpha^I$ and $q_Ih^I$ (as we will see in section \ref{F0}). Thus, due to holomorphy, the quantum correction to $\F_\sg$ should be a function of $q_I Z^I$. From shift symmetries, it actually should be a a function of $e^{2\pi i q_I Z^I}$. Thus we conclude that the general form of the contribution of one BPS multiplet to $\F_\sg$, which we will usually denote by $\F_\sg^{\rm q}$, is:
 \begin{equation}
 \label{shiftF}
 \F_\sg^{\rm q} = (X^0)^{2-2\sg}\sum_{k> 0} c_{k,\sg} e^{2\pi i k q_I Z^I},
 \end{equation}
where we did not allow negative values of $k$, as the contribution $\propto e^{-2\pi k M}$, $M=q_I h^I>0$ should decay faster for more massive particles, rather than exponentially grow (if we had $q_Ih^I<0$, this would describe an antiparticle, and only negative $k$ would have to be present in the sum, for the same reason).

\subsection{Constraints on $\F_\sg$ from parity}\label{parity}
M-theory has a discrete symmetry which is often called ``parity'' and is a combination of some orientation reversing diffeomorphism in 11d and a sign change of the 3-form gauge field $C$. This symmetry descends in an obvious way to the symmetry of the 5d action (\ref{act5d}), and then to 4 dimensions as well. The fields $A^I_\mu$ and $\alpha^I$, which originate from the 11d $C$-field, get an extra minus sign, while the field $A^0_\mu$, which is a Kaluza-Klein gauge field, does not. So, to summarize, the 4d supergravity we obtain should be invariant under the parity defined as an orientation reversal combined with the following:
\begin{align}
A^I_\mu &\to -A^I_\mu,\cr
A^0_\mu &\to A^0_\mu,\cr
{\rm Re}\, Z^I\equiv \alpha^I &\to -\alpha^I.
\end{align}

How does it constrain the form of $\F_\sg$? Since $d=4, N=2$ supergravity written in a given metric frame lifts in a unique way to the conformal supergravity, the parity symmetry also lifts there. It can then be extended to the symmetry of the superspace action. Since parity switches chiralities, we can conclude that the two terms of the form:
\begin{equation}
-i\int\d^4\theta\, \F_\sg(\X)\W^{2\sg} + i\int\d^4\bar\theta\, \bar\F_\sg(\bar\X)\bar\W^{2\sg}
\end{equation}
are switched by parity, where the second term is the complex conjugate of the first and $\bar\theta$ are superspace coordinates of opposite chirality. This means, in particular, that for all $\g\ge 0$, $-i\F_\sg(X)$ goes into $i\bar\F_\sg(\bar{X})$ under parity. We are working in the gauge where $iX^0>0$, and so if we consider the non-homogeneous function $\hat{\F}_\sg(Z)=(X^0)^{2\sg-2}\F_\sg(X)$, we also find that parity complex conjugates $i\hat{\F}_\sg(Z)$.

Now, since parity multiplies $\alpha^I$ by $-1$, it means that all terms $\hat{\F}_\sg(Z)$ in the GV formula go to $-\bar{\hat{\F}_\sg(Z)}$ under $\alpha^I\to-\alpha^I$. This implies that they should be imaginary at $\alpha^I=0$.\footnote{For analytic $\F_\sg(Z)$, these two conditions are actually equivalent.} In particular, $c_{k,\sg}$ in (\ref{shiftF}) are imaginary. We will use it soon.

\section{Computation of $\F_0$}\label{F0}
Now we consider a light massive hypermultiplet coupled to the 5d supergravity (\ref{act5d}) which has enough scalars $h^I$ (we will explain this requirement in Section \ref{val}). For the purposes of one-loop computation, the global geometry of space parametrized by scalars in the hypermultiplet is irrelevant. So, this multiplet can be described by a pair of complex scalars $z^i,~i=1,2$ and a Dirac spinor $\Psi$ in 5d. The quadratic action on the flat background with no gauge fields turned on is:
\begin{equation}
\label{hypact}
S_h = \int \d^5x \left(\sum_{i=1}^2 (-|\partial z^i|^2 - M^2 |z^i|^2) +\bar\Psi^c\slashed{\partial}\Psi - M\bar\Psi^c \Psi \right).
\end{equation}

We want to determine its contribution to the term $\F_0$ in the 4d effective superpotential. Our strategy is to determine first its contribution to the 4d Kahler metric on the vector multiplets moduli space, and then, since this metric is encoded in $\F_0$, to reconstruct the hypermultiplet contribution to $\F_0$.

To find the contribution to the Kahler metric, we need to compute the effective action governing fluctuations of vector multiplet scalars on the flat background $\R^{3,1}\times S^1$, which is the simplest possible background consistent with our problem.

Let us describe the precise setup. Note first that the expected answer has a known form (\ref{shiftF}), in which we only have to determine constants $c_{k,\sg}$. To do that, we can choose any convenient values for the background fields. One such field is the radius of the M-theory circle $e^\sigma$. Even though conceptually the computation happens at the large radius (as explained in \cite{GVrev}), the field theory computation is valid at the arbitrary radius once we have the action (\ref{hypact}) and know what to compute. For simplicity, we assume that the radius of $S^1$ is constant and is equal to $1$, that is $e^\sigma=1$. We also do not switch on holonomies, $\alpha^I=0$. We allow the 5d scalars $h^I$ to depend on the point of $\R^{3,1}$, while they still should be invariant under translations along $S^1$. Since the mass of the BPS particle in 5d is expressed through its charges $q_I$ in the following way (see \cite{GVrev}):
\begin{equation}
M=\sum_I q_I h^I,
\end{equation}
$M(x)$ is allowed to fluctuate around its constant background value $M$, with fluctuations depending only on the point of $\R^{3,1}$. Now, to determine the Kahler metric deformation, we need to find a term in the effective action which is quadratic in $M(x)$ and has precisely two derivatives. Since the effective action is $S_{eff}=-i\ln\int \mathcal{D}z^i\mathcal{D}\Psi e^{iS_h}$, and we are looking for
\begin{equation}
\frac{\delta^2 S_{eff}}{\delta M(x)\delta M(y)}\Big|_{M=const},
\end{equation}
it is clear that all we need to compute is a connected two-point function of the mass terms:
\begin{equation}
\label{twomass}
-i\left\langle \left(2M\sum_i |z^i|^2 + \bar\Psi^c\Psi\right)\left(2M\sum_i |z^i|^2 + \bar\Psi^c\Psi\right) \right\rangle_{\rm conn},
\end{equation}
and then, in the momentum space representation with an external momentum $p$, to extract the $p^2$-part of the answer. This will give the one-loop Kahler metric deformation due to the light hypermultipet.

After we calculate the Kahler metric deformation, we will have to reconstruct the prepotential deformation from it. We use notation ${\rm cl}$ and ${\rm q}$ to distinguish classical and one-loop parts, so for example the full prepotential is $\F_0 = \F_0^{\rm cl} + \F_0^{\rm q}$, were the classical part is given by (\ref{class}). The Kahler metric deformation is written in terms of the scalars $Z^I=X^I/X^0$ of Poincare supergravity. However, $\F_0(X)$ is a function of conformal scalars $X^\Lambda$, so to reconstruct it, we should know the expression of $X^0$ in terms of $Z^I$ and $\bar{Z}^I$. Reconstructing $\F_0$ includes some subtleties, which we will discuss in detail later, after the two-point function computation.

\subsection{The two-point function computation}
We proceed to compute (\ref{twomass}) here. First of all, we need to know the relevant Green's functions on $\R^{3,1}\times S^1$. Let $x^\mu$ be coordinates on $\R^{3,1}$ and $y\in[0,2\pi]$ be a coordinate on $S^1$. If $G_0(x,y)$ and $D_0(x,y)$ are the Green's functions for bosons and fermions respectively on $\R^{4,1}$, i.e. they satisfy:
\begin{align}
(\partial^2 - M^2)G_0(x,y)&=\delta^{(4)}(x)\delta(y),\cr
(\slashed{\partial}-M)D_0(x,y)&=\delta^{(4)}(x)\delta(y),
\end{align}
then the Green's functions on $\R^{3,1}\times S^1$ are just:
\begin{align}
\label{ksum}
G(x,y)&=\sum_{k\in \Z}G_0(x,y+2\pi k),\cr
D(x,y)&=\sum_{k\in\Z}D_0(x,y+2\pi k).
\end{align}
Then (\ref{twomass}) gives:
\begin{equation}
\label{step1}
-8iM^2 G(x_1-x_2,y_1-y_2)G(x_2-x_1,y_2-y_1) + i \Tr\big[D(x_1-x_2,y_1-y_2)D(x_2-x_1,y_2-y_1)\big],
\end{equation}
where $(x_1,y_1)$ and $(x_2,y_2)$ are the space-time points where the two mass terms are inserted. If $\mathcal{K}(x_1,y_1;x_2,y_2)$ is the expression (\ref{step1}), then the term in the effective action is 
\begin{equation}
\int \d^4x_1\d y_1\d^4x_2\d y_2 \,\mathcal{K}(x_1,y_1;x_2,y_2) M(x_1)M(x_2). 
\end{equation}
We note that since $M(x)$ is independent of the circle direction $y$, we can integrate (\ref{step1}) over $y_1$ and $y_2$, or over $y\equiv y_1-y_2$ and $y_2$. Another obvious step is to pass to the momentum representation for the $\R^{3,1}$ directions. Now we have
\begin{equation}
\int \d y_1 \d y_2\,\mathcal{K}(x_1,y_1;x_2,y_2) = \int \frac{\d^4 p}{(2\pi)^4}\,\mathcal{K}(p)e^{i p(x_1-x_2)},
\end{equation}
and this $\mathcal{K}(p)$ is given by
\begin{equation}
\mathcal{K}(p) = -2\pi i\int_0^{2\pi}\d y\int\frac{\d^4q}{(2\pi)^4}\Big(8M^2 G(q,y)G(q-p,-y) - \Tr\big[D(q,y)D(q-p,-y)\big]\Big).
\end{equation}
If we substitute (\ref{ksum}), this becomes: 
\begin{align}
\mathcal{K}(p) = -2\pi i\sum_{k_1,k_2}\int_0^{2\pi}\d y\int\frac{\d^4q}{(2\pi)^4}\Big(8M^2 G_0(q,y-2\pi k_1)G_0(q-p,-y - 2\pi k_2)\cr
 - \Tr\big[D_0(q,y-2\pi k_1)D_0(q-p,-y-2\pi k_2)\big]\Big).
\end{align}
This quantity is represented by the Feynman diagram on Figure 1, where scalars and bosons run inside the loop, and we label internal lines of the loop by the corresponding 4d momentum and the winding number $k$.
\begin{figure}[h]
\caption{The two-point function of mass terms. Internal lines are labeled by the 4d momentum and the winding number.}
\centering
\includegraphics[scale=1]{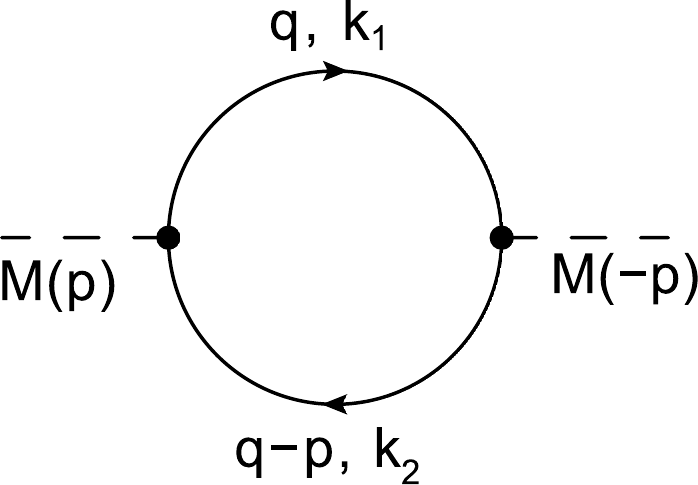}
\end{figure}
It is clear from the picture that $k_1+k_2$ plays the role of the total winding number of the particle as it circles the loop in the diagram. Another way to see it is to reintroduce non-zero constant holonomies $\alpha^I$. These would just shift the momentum in the circle direction by $w \to w + q_I\alpha^I$ and contribute an overall factor $e^{-iq_I\alpha^I y}$ both in $G_0(p,y)$ and $D_0(p,y)$. Then, in the above expression for $\mathcal{K}(p)$, the only effect of holonomies would be to introduce an overall factor $e^{2\pi i (k_1+k_2)q_I\alpha^I}$, thus showing that $k_1+k_2$ is indeed interpreted as the total winding number of the loop.

We need explicit expressions for $G_0$ and $D_0$ in a ``mixed'' representation, where momentum is used for the $x^\mu$ directions and position coordinate is used for the $y$ direction. It is easy to find that:
\begin{align}
D_0(p,y)&=\int_{-\infty}^\infty \frac{\d w}{2\pi} e^{i w y} \frac{M-i\slashed{p}-iw\Gamma^5}{p^2 + w^2 + M^2}=\frac{M-i\slashed{p}}{2\sqrt{p^2 + M^2}}e^{-|y|\sqrt{p^2 + M^2}} + \frac{{\rm Sign}(y)}{2}\Gamma^5e^{-|y|\sqrt{p^2 + M^2}},\cr
G_0(p,y)&=\int_{-\infty}^\infty \frac{\d w}{2\pi} e^{i w y}\frac{1}{p^2 + w^2 + M^2}=\frac{e^{-|y|\sqrt{p^2 + M^2}}}{2\sqrt{p^2 + M^2}}.
\end{align}
Substituting this into our expression for $\mathcal{K}(p)$, computing traces of gamma matrices and considering a given fixed $k=k_1+k_2$, we get
\begin{align}
\label{twp}
&-2\pi i\sum_{k_1+k_2=k}\int_0^{2\pi}\d y\int\frac{\d^4q}{(2\pi)^4}\Big(\frac{M^2+q^2-pq}{\sqrt{q^2+M^2}\sqrt{(q-p)^2+M^2}}\cr
&+{\rm Sign}(y-2\pi k_1){\rm Sign}(y+2\pi k_2) \Big)e^{-|y-2\pi k_1|\sqrt{q^2+M^2}-|y+2\pi k_2|\sqrt{(q-p)^2+M^2}}.
\end{align}
For this computation and for the computation in the next section, we need the following two formulas:
\begin{align}
\label{AB}
\int_0^{2\pi}\d y\sum_{k_1+k_2=k}e^{-|y-2\pi k_1|A - |y+2\pi k_2|B}=\frac{e^{-2\pi |k| B}-e^{-2\pi |k|A}}{A-B} + \frac{e^{-2\pi |k|A}+e^{-2\pi |k|B}}{A+B},
\end{align}
\begin{align}
\label{ABSign}
\int_0^{2\pi}\d y\sum_{k_1+k_2=k}e^{-|y-2\pi k_1|A - |y+2\pi k_2|B}{\rm Sign}(y-2\pi k_1){\rm Sign}(y+2\pi k_2)\cr
=-\frac{e^{-2\pi |k| B}-e^{-2\pi |k|A}}{A-B} + \frac{e^{-2\pi |k|A}+e^{-2\pi |k|B}}{A+B}.
\end{align}

Applying them to (\ref{twp}), we get:
\begin{align}
\label{d4qhuge}
&-2\pi i\int\frac{\d^4q}{(2\pi)^4}\Bigg[\frac{M^2 + q^2 - pq}{\sqrt{q^2+M^2}\sqrt{(q-p)^2+M^2}}\times\cr
&\left(\frac{e^{-2\pi |k| \sqrt{(q-p)^2+M^2}}-e^{-2\pi |k|\sqrt{q^2+M^2}}}{\sqrt{q^2+M^2}-\sqrt{(q-p)^2+M^2}} + \frac{e^{-2\pi |k|\sqrt{q^2+M^2}}+e^{-2\pi |k|\sqrt{(q-p)^2+M^2}}}{\sqrt{q^2+M^2}+\sqrt{(q-p)^2+M^2}} \right)\cr
&-\frac{e^{-2\pi |k| \sqrt{(q-p)^2+M^2}}-e^{-2\pi |k|\sqrt{q^2+M^2}}}{\sqrt{q^2+M^2}-\sqrt{(q-p)^2+M^2}} + \frac{e^{-2\pi |k|\sqrt{q^2+M^2}}+e^{-2\pi |k|\sqrt{(q-p)^2+M^2}}}{\sqrt{q^2+M^2}+\sqrt{(q-p)^2+M^2}}\Bigg].
\end{align}

This expression is perfectly convergent for $k\ne 0$ and we are going to compute it shortly, but first let us say a few words about $k=0$. 
\subsubsection*{\emph{A digression about} $k=0$}

The case $k=0$ corresponds, in the particle language, to the contribution of closed trajectories that do not have any net winding number. Such trajectories in $\R^4\times S^1$ can be lifted to closed trajectories in $\R^5$. Thus the $k=0$ term should be understood as a contribution to the 5d effective action. It then contributes to the 4d effective action through the classical dimensional reduction. As was explained in \cite{GVrev}, only two F-terms can receive contributions from the classical dimensional reduction. Those are precisely the $\F_0$ and $\F_1$ that are being studied in this paper. The $\F_1$ term will be discussed in the next section, while for the prepotential $\F_0$, the only possible contributions from dimensional reduction originate from the 5d action (for supergravity with vector multiplets) with no more than 2 derivatives. Such an action in 5d is completely fixed by supersymmetry in terms of the coefficients $\CC_{IJK}$ (see \cite{jord}). Dimensional reduction then gives the prepotential (\ref{class}) in 4d depending on these coefficients. So the only possibility for the $k=0$ contribution to affect the $\F_0$ term in 4d is to shift the values of $\CC_{IJK}$ in the 5d effective action. This does not happen. One way to see it is to note that the 5d action has a Chern-Simons term $C_{IJK}V^I\wedge \d V^J\wedge \d V^K$. It gives rise to the term $C_{IJK}\alpha^I F^J\wedge F^J$ in the 4d action, where $\alpha^I=V^I_y$ are holonomies along the circle. Any quantum computation will depend on holonomies through the combination $e^{2\pi i \alpha^I}$, and thus the term $C_{IJK}\alpha^I F^J\wedge F^J$ cannot be shifted. \footnote{From the string theory side, the values of $\CC_{IJK}$ are given by the string three-point amplitudes on a sphere $S^2$ with one insertion of the NS-NS vertex operator corresponding to the scalar $\alpha^I$ and two inservions from the R-R sector corresponding to the field strengths $F^I_{\mu\nu}$ (see \cite{geomtype2}).}.

\subsubsection*{\emph{Back to the computation}}

Now, for $k\ne 0$, we want to Taylor expand the integrand in (\ref{d4qhuge}) and pick out the $p^2$-term in the expansion. Schematically, there will be two kinds of terms:
\begin{equation}
\int \frac{\d^4q}{(2\pi)^4}\left[f_1(q^2)p^2 + f_2(q^2) (pq)^2 \right].
\end{equation}
In this type of integral one usually performs a Wick rotation $q^0=-iq^4$, and then notes that, due to the spherical symmetry, $q_\mu q_\nu$ can be replaced by $\frac{q^2}{4}\eta_{\mu\nu}$. After that, we have:
\begin{equation}
-i\int\frac{\d^4q_E}{(2\pi)^4}\left[f_1(q^2_E)+f_2(q^2_E)q^2_E/4\right]p_E^2.
\end{equation}
By going to spherical coordinates and recalling that the volume of the unit $3$-sphere is $2\pi^2$, one has to compute:
\begin{equation}
-\frac{i\pi}{(2\pi)^3}\int_0^\infty q^3\d q\left[f_1(q^2) + f_2(q^2)q^2/4\right]p^2.
\end{equation}
By applying this to (\ref{d4qhuge}) (after Taylor expansion), we get the following expression at the $p^2$-order:
\begin{align}
-\frac{\pi}{(2\pi)^2}p^2\int_0^\infty q^3\d q\Bigg[ \frac{e^{-2\pi |k|\sqrt{M^2+q^2}} q^2 \left(3+4\pi^2 k^2 M^2+4\pi^2k^2q^2+6\pi |k|\sqrt{M^2+q^2}\right)}{8 \left(M^2+q^2\right)^{5/2}}\cr 
-\frac{e^{-2\pi |k|\sqrt{M^2+q^2}}}{\left(M^2+q^2\right)^{3/2}}-\frac{2\pi |k| e^{-2\pi |k|\sqrt{M^2+q^2}} }{M^2+q^2}\Bigg].
\end{align}
By an obvious change of variables $x=\sqrt{M^2+q^2}$, this is transformed into:
\begin{align}
-\frac{\pi}{(2\pi)^2}p^2\int_M^\infty\d x\, x(x^2-M^2)\Bigg[\frac{e^{-2 |k| \pi  x} \left(x^2-M^2\right) \left(3+4\pi^2k^2M^2+6\pi |k|  x+4\pi^2k^2\left(x^2-M^2\right)\right)}{8 x^5}\cr
-\frac{2\pi |k| e^{-2\pi |k| x} }{x^2}-\frac{e^{-2\pi |k| x}}{x^3}\Bigg],
\end{align}
which gives:
\begin{align}
\frac{\pi e^{-2\pi |k|M}}{(2\pi)^3|k|}p^2.
\end{align}
We sum this over $k\ne 0$ ($k$ and $-k$ pair up) and get the corresponding kinetic term deformation in coordinate space: 
\begin{equation}
\label{kahdef}
-\frac{1}{2}\sum_{k=1}^\infty \frac{ e^{-2\pi k M}}{(2\pi)^2k}\partial_\mu M(x) \partial^\mu M(x) = -\frac{1}{2}\sum_{k=1}^\infty \frac{ e^{-2\pi k M}}{(2\pi)^2k} q_I q_J \partial_\mu h^I \partial^\mu h^J.
\end{equation}
\subsection{Reconstructing $\F_0$}
Now we aim to reconstruct the expression for $\F_0$ from the Kahler metric deformation we have computed. An important observation one should make first is that the one-loop quantum corrections also include contributions to the effective action that describe couplings of the vector multiplets scalars $Z^I$ to the scalar curvature $R$. That is, effective supergravity emerges written in a non-Einstein frame. If we denote the corresponding one-loop contribution as $\frac{1}{2}\phi(Z,\bar{Z})R$, then the part of the Lagrangian density including also kinetic energy of scalars, written at the point with zero holonomies $\alpha^I=0$, is:
\begin{equation}
\label{oneloopact}
\frac{1}{2}(1+\phi(Z,\bar{Z}))R + \frac{3}{2}\CC_{IJK}h^I\partial_\mu h^J\partial^\mu h^K -\frac{1}{2}\sum_{k=1}^\infty \frac{ e^{-2\pi k M}}{(2\pi)^2k} q_I q_J \partial_\mu h^I \partial^\mu h^J.
\end{equation}
We could find this $\phi(Z,\bar{Z})$ by similarly computing the two-point function of some scalar $Z^I$ with the metric. However, there is no need to do it as the structure of $N=2$ supergravity determines this function in terms of the quantities we have already calculated, as we will see soon.

We want to compare the deformed metric on scalars in (\ref{oneloopact}) with the formulas from the Section \ref{sug4d}, namely with the general expression for the Kahler metric (\ref{kinstand}) in the Einstein frame. However, since the action (\ref{oneloopact}) is written in a non-Einstein frame, we have to rescale metric first, writing the action in the Einstein frame:
\begin{equation}
\frac{1}{2}R + \frac{3}{2}(1+\phi(Z,\bar{Z}))^{-1}\CC_{IJK}h^I\partial_\mu h^J\partial^\mu h^K -\frac{1}{2}(1+\phi(Z,\bar{Z}))^{-1}\sum_{k=1}^\infty \frac{ e^{-2\pi k M}}{(2\pi)^2k} q_I q_J \partial_\mu h^I \partial^\mu h^J.
\end{equation}
Keeping only the first order corrections, we find:
\begin{equation}
\label{computed}
-\frac{3}{2}\phi(Z,\bar{Z})\CC_{IJK}h^I\partial_\mu h^J\partial^\mu h^K -\frac{1}{2}\sum_{k=1}^\infty \frac{ e^{-2\pi k M}}{(2\pi)^2k} q_I q_J \partial_\mu h^I \partial^\mu h^J,
\end{equation}
which is the desired Kahler metric deformation. We want to compare it with the deformation of (\ref{kinstand}) under $\F_0 = \F_0^{\rm cl} + \F_0^{\rm q}$, where $\F_0^{\rm cl}$ is the classical prepotential (\ref{class}). Such a prepotential deformation results in $N_{\Lambda\Sigma}=N_{\Lambda\Sigma}^{\rm cl}+N_{\Lambda\Sigma}^{\rm q}$ and, through the gauge condition $N_{\Lambda\Sigma}X^\Lambda\bar{X}^\Sigma=-1$, in the deformation of the expression for $X^0$ in terms of other scalars. It is straightforward to find the first order correction of (\ref{kinstand}) at $\alpha^I=0$ and $e^\sigma=1$:
\begin{equation}
\frac{1}{4}N_{IJ}^{\rm q}\partial_\mu h^I\partial^\mu h^J + \frac{1}{4}(N_{IJ}^{\rm q}h^I h^J+N_{00}^{\rm q}) \frac{3}{2}\CC_{IJK}h^I\partial_\mu h^J\partial^\mu h^K.
\end{equation}
Recalling the general expression for $\F_0^{\rm q}$ (\ref{shiftF}) deduced from shift symmetries, one can further write this as:
\begin{align}
\label{predicted}
&-2\pi^2 \partial_\mu M\partial^\mu M\sum_{k>0}k^2 {\rm Im}\, (c_{k,0}) e^{-2\pi k M}\cr 
&+ \left( 2\pi M \sum_{k>0}k\,{\rm Im}\, (c_{k,0})e^{-2\pi k M} + \sum_{k>0}{\rm Im}\, (c_{k,0})e^{-2\pi k M}  \right) \frac{3}{2}\CC_{IJK}h^I\partial_\mu h^J\partial^\mu h^K,
\end{align}
where we used $M=q_I h^I$. We now want to equate this to the result of the one-loop calculation given in (\ref{computed}). Also, it is useful to realize that at $\alpha^I=0$, the function $\phi(Z,\bar{Z})$ is really a function $\phi(M)$ of $M=q_I h^I$ only, simply because it is a one-loop effect due to the particle of mass $M$. Equating (\ref{computed}) with (\ref{predicted}) and slightly rearranging, we get:
\begin{align}
\label{eq_c_phi}
&2\pi^2 \partial_\mu M\partial^\mu M\sum_{k>0}k^2 \left( {\rm Im}\, (c_{k,0}) - \frac{1}{(2\pi)^4k^3} \right) e^{-2\pi k M} = \cr 
&=\left( \phi(M)+ 2\pi M \sum_{k>0}k\,{\rm Im}\, (c_{k,0})e^{-2\pi k M} + \sum_{k>0}{\rm Im}\, (c_{k,0})e^{-2\pi k M}  \right) \frac{3}{2}\CC_{IJK}h^I\partial_\mu h^J\partial^\mu h^K,\cr
\end{align}
which is the equation for the unknown coefficients $c_{k,0}$ and the unknown function $\phi(M)$. When written in such a way and if there are enough scalars $h^I$ in the theory, one can show\footnote{\label{variation}If there are enough scalars, one can find such a constant (i.e. independent of the space-time point) infinitesimal variation $\delta h^I$ that $\CC_{IJK}\delta h^I \partial_\mu h^J\partial^\mu h^K$ is non-zero, while $\delta M=q_I \delta h^I=0$. Of course, constraint $\CC_{IJK}h^Ih^Jh^K=1$ defining the hypersurface $\mathcal{M}_h$ should be preserved too. Under such a variation in $h^I$, the equation (\ref{eq_c_phi}) should be preserved. But since $\delta M=0$, the only term whose variation is non-zero is $\CC_{IJK} h^I \partial_\mu h^J\partial^\mu h^K$. Thus the expression in parenthesis by which it is multiplied should be zero for the equation to hold, which immediately implies (\ref{ans_c_ph}). There are $b_2(Y)$ scalars $h^I, I=1\dots b_2(Y)$, where $b_2(Y)$ is a second Betti number of $Y$. To have ``enough scalars'', we can take $b_2(Y)\ge 4$. To show this, put $\partial_\mu h^I= a_{\mu}\delta h^I$, i.e. assume that the gradient is parallel to the variation that we are seeking with some proportionality factor $a_\mu$ such that $a_\mu a^\mu\ne 0$ (we can obviously do that). The fact that $\CC_{IJK}h^Ih^Jh^K=1$ is preserved means that $\delta h^I$ is tangent to $\mathcal{M}_h$. Also, as mentioned above, we have $q_I \delta h^I=0$. Also, we want $\CC_{IJK}\delta h^I \partial_\mu h^J\partial^\mu h^K= a_\mu a^\mu \CC_{IJK}\delta h^I \delta h^J\delta h^K\ne 0$. When $b_2(Y)\ge 4$, the tangent space to $\mathcal{M}_h$ is at least three-dimensional, and $q_I \delta h^I=0$ gives a subspace of dimension at least two. In such a space, we can clearly find such $\delta h^I$ that a single condition $\CC_{IJK}\delta h^I \delta h^J\delta h^K\ne 0$ is satisfied, and this is the variation we need, so $b_2(Y)\ge 4$ is enough. However, in Subsection \ref{val} we will explain that the answer we get is valid for any $b_2(Y)$.} that the only possible way to satisfy it is to set both sides to zero. Recalling that $c_{k,0}$ are imaginary due to parity, this gives:
\begin{align}
\label{ans_c_ph}
c_{k,0}&=\frac{i}{(2\pi)^4k^3},\cr
\phi(M)&=-\sum_{k>0}\frac{M}{(2\pi)^3k^2}e^{-2\pi k M} - \sum_{k>0}\frac{1}{(2\pi)^4k^3}e^{-2\pi k M}.
\end{align}
With such values of $c_{k,0}$, we get:
\begin{equation}
\label{f0result}
\F^{\rm q}_0 = \frac{i}{(2\pi)^4}(X^0)^2\sum_{k=1}^\infty \frac{1}{k^3}e^{2\pi i k q_I Z^I},
\end{equation}
which agrees with the GV formula as claimed in \cite{GVrev}. It is now also obvious that for $\alpha^I\ne 0$, the expression for $\phi(Z,\bar{Z})$ is:
\begin{equation}
\label{phigen}
\phi(Z,\bar{Z})=-N_{\Lambda\Sigma}^{\rm q}X^\Lambda\bar{X}^\Sigma=-\frac{1}{4}e^{-3\sigma}N_{\Lambda\Sigma}^{\rm q}Z^\Lambda\bar{Z}^\Sigma.
\end{equation}
\subsubsection{The case of arbitrary $b_2(Y)$}\label{val}
In the derivation of (\ref{ans_c_ph}), we used the assumption that there are enough scalars, namely that $b_2(Y)\ge 4$, as explained in the footnote \ref{variation}. However, there exist Calabi-Yau spaces with $b_2(Y)<4$. For example, the quintic threefold has $b_2(Y)=1$, which is the minimal possible value. In fact, the case of $b_2(Y)=1$ seems to be even more problematic, because the 5d theory obtained by compactification on such a manifold has no vector multiplets and so no corresponding scalars. But our approach was to compute the Kahler metric for those scalars, so their existence was essential.

A  possible way around is that the formula (\ref{f0result}), describing the contribution of a single 5d hypermultiplet to $\F_0$, is universal and holds for any $b_2(Y)$. Once we know that the corresponding 5d BPS multiplet exists, this formula gives the answer irrespective of how big or small $b_2(Y)$ is. For $b_2(Y)\ge 4$, this already follows from our derivation, but for the cases of small $b_2(Y)$, one has to give a separate argument.

To do this, notice that we could set up a different computation of $\F_0$. Namely, we could use the kinetic energy of gauge fields. It has two good properties. One is that it is Weyl-invariant, so rescaling the metric into the Einstein frame would not affect the one-loop deformation of the kinetic term (unlike it was for scalars in (\ref{oneloopact})-(\ref{computed})). Another is that the matrix of couplings (\ref{Ncal}) does not depend on the dilatational gauge, i.e. on the expression for $X^0$, so that the gauge fields kinetic term deformation is directly related to $N_{\Lambda\Sigma}^{\rm q}$. So we could just compute the two-point function of 5d gauge fields (they exist for all $b_2(Y)$, unlike scalars), and get $\F_0^{\rm q}$ out of it directly. A disadvantage of such an approach is that it seems to be much more technically involved than what we have done here, and one would also need to know how to couple the minimal action (\ref{hypact}) to gauge fields in a proper supersymmetric way. That is why we have chosen scalars for the computation. But it would clearly depend only on the properties of the 5d hypermultiplet, and not on $b_2(Y)$. Existence of such an alternative computation establishes our claim that (\ref{f0result}) provides a universal answer.

\section{Computation of $\F_1$}\label{F1}
Now we consider the same light hypermultiplet as in (\ref{hypact}), but here we determine its contribution to $\F_1$. The term $\F_1$ gives rise to a variety of interactions in the 4d effective action, and every one of them can potentially be used to set up a computation of $\F_1$. We find the following term:
\begin{equation}
({\rm Im}\,\F_1)R^2\equiv ({\rm Im}\,\F_1)R_{\mu\nu\lambda\rho}R^{\mu\nu\lambda\rho}
\end{equation}
to be the most useful for this purpose. This term can be understood as a response to a small metric perturbation. Thus, it can be computed from the two-point function of the symmetric stress-energy tensor of the action (\ref{hypact}). We consider a small metric perturbation around the flat 4d Minkowski background (the $\R^{3,1}$ part of $\R^{3,1}\times S^1$):
\begin{equation}
g_{\mu\nu}=\eta_{\mu\nu}+h_{\mu\nu},
\end{equation}
and we assume no metric perturbations in the circle direction. That is, the metric remains $\d s^2= g_{\mu\nu}\d x^\mu \d x^\nu + dy^2$. With the TT-gauge condition:
\begin{align}
h_\mu^\mu&=0\cr
\partial_\mu h^{\mu\nu}&=0,
\end{align}
we have: $R^2 = \partial_\lambda\partial_\sigma h^{\mu\nu}\partial^\lambda\partial^\sigma h_{\mu\nu} + O(h^3)$. So we will compute the following interaction:
\begin{equation}
8({\rm Im}\,\F_1)h^{\mu\nu}(\partial^2)^2h_{\mu\nu}.
\end{equation}
If $T_{MN}$ is a symmetric stress-energy tensor of (\ref{hypact}), then for small perturbations $h_{\mu\nu}$ of the metric, the leading order contribution to $({\rm Im}\,\F_1) R^2$ at one loop comes from:
\begin{equation}
\frac{1}{4}\int \d^5x\d^5y \langle T_{\mu\nu}(x)T_{\lambda\rho}(y)\rangle h^{\mu\nu}(x)h^{\lambda\rho}(y).
\end{equation}
The useful relation to extract the one-loop answer $\F_1^{\rm q}$ is:
\begin{equation}
\label{tpf}
\int\d^5 x_1\d^5 x_2\langle T_{\mu\nu}(x_1)T_{\lambda\rho}(x_2)\rangle h^{\mu\nu}(x_1)h^{\lambda\rho}(x_2) = -64i\int\d^4 x ({\rm Im}\,\F_1^{\rm q}) h^{\mu\nu}(\partial^2)^2 h_{\mu\nu} +\dots,
\end{equation}
where the ellipsis stands for terms with the wrong number of derivatives.
\subsection{The two-point function computation}
The symmetric stress-energy tensor is
\begin{equation}
T_{\mu\nu}=-2\sum_i \bar{z}^i\left(\overleftarrow\partial_{(\mu}\overrightarrow\partial_{\nu)} \right)z^i + \frac{1}{2}\bar\Psi^c \left(\gamma_{(\mu}\overrightarrow\partial_{\nu)} - \gamma_{(\mu}\overleftarrow\partial_{\nu)} \right)\Psi - \eta_{\mu\nu}\mathcal{L}.
\end{equation}
Formula (\ref{tpf}) implies that, due to the tracelessness of $h_{\mu\nu}$, the $\eta_{\mu\nu}\mathcal{L}$ term in the expression for $T_{\mu\nu}$ is unimportant.

Since the leading contribution to $R_{\mu\nu\lambda\rho}R^{\mu\nu\lambda\rho}$ is proportional to $(\Box h)^2$, we need to find the $(p^2)^2$-order term of the $\langle T_{\mu\nu}T_{\lambda\rho}\rangle$ two-point function. We identify the contribution of bosons first:
\begin{equation}
8\times 2\pi\int_0^{2\pi}\d y\int\frac{\d^4p}{(2\pi)^4} h^{\mu\nu}(-p)h^{\lambda\rho}(p)\int\frac{d^4q}{(2\pi)^4}(q-p)_\mu q_\nu q_\lambda (q-p)_\rho G(q,y)G(q-p,-y).
\end{equation}
Because of $\partial_\mu h^{\mu\nu}=0$, we have $p^\mu h_{\mu\nu}(p)=0$, and so the important part is:
\begin{equation}
8\times 2\pi\int_0^{2\pi}\d y\int\frac{\d^4p}{(2\pi)^4} h^{\mu\nu}(-p)h^{\lambda\rho}(p)\int\frac{d^4q}{(2\pi)^4}q_\mu q_\nu q_\lambda q_\rho G(q,y)G(q-p,-y).
\end{equation}
The contribution of fermions is:
\begin{equation}
\frac{1}{4}\int\frac{\d^4p}{(2\pi)^4} h^{\mu\nu}(-p)h^{\lambda\rho}(p)\int\frac{d^4q}{(2\pi)^4}{\rm Tr}\left\{(\gamma_{(\mu}q_{\nu)} - \gamma_{(\mu}(p-q)_{\nu)})D(q,y)(\gamma_{(\lambda}(q-p)_{\rho)} + \gamma_{(\lambda}q_{\rho)})D(q-p,-y) \right\},
\end{equation}
And for the same reason, $p^\mu h_{\mu\nu}=0$, the relevant part is:
\begin{equation}
\int\frac{\d^4p}{(2\pi)^4} h^{\mu\nu}(-p)h^{\lambda\rho}(p)\int\frac{d^4q}{(2\pi)^4}{\rm Tr}\left\{\gamma_{\mu}q_{\nu}D(q,y)\gamma_{\lambda}q_{\rho}D(q-p,-y) \right\}.
\end{equation}
So, we have to find the $(p^2)^2$-term of this expression:
\begin{align}
2\pi\int_0^{2\pi}\d y\int\frac{d^4q}{(2\pi)^4}\left[8q_\mu q_\nu q_\lambda q_\rho G(q,y)G(q-p,-y) + {\rm Tr}\left\{\gamma_{(\mu}q_{\nu)}D(q,y)\gamma_{(\lambda}q_{\rho)}D(q-p,-y) \right\}\right].
\end{align}
The following steps are as in the $\F_0$ case. We have:
\begin{align}
2\pi\sum_{k_1,k_2}\int_0^{2\pi}\d y\int\frac{d^4q}{(2\pi)^4}\Big[8q_\mu q_\nu q_\lambda q_\rho G_0(q,y-2\pi k_1)G_0(q-p,-y-2\pi k_2) \cr
+ {\rm Tr}\left\{\gamma_{(\mu}q_{\nu)}D_0(q,y-2\pi k_1)\gamma_{(\lambda}q_{\rho)}D_0(q-p,-y-2\pi k_2) \right\}\Big],
\end{align}
and for given $k_1+k_2=k$, we get:
\begin{align}
&2\pi\sum_{k_1+k_2=k}\int_0^{2\pi}\d y\int\frac{d^4q}{(2\pi)^4}\Bigg[\frac{2q_\mu q_\nu q_\lambda q_\rho}{\sqrt{q^2+M^2}\sqrt{(q-p)^2+M^2}}\cr
&+\frac{M^2 q_{(\nu}g_{\mu)(\lambda}q_{\rho)}}{\sqrt{q^2+M^2}\sqrt{(q-p)^2+M^2}}-\frac{q_\mu q_\nu (q-p)_{(\lambda}q_{\rho)} + (q-p)_{(\mu}q_{\nu)}q_\lambda q_\rho - q(q-p) q_{(\nu}g_{\mu)(\lambda}q_{\rho)}}{\sqrt{q^2+M^2}\sqrt{(q-p)^2+M^2}}\cr
&+q_{(\nu}g_{\mu)(\lambda}q_{\rho)} {\rm Sign}(y-2\pi k_1){\rm Sign}(y+2\pi k_2)\Bigg]e^{-|y-2\pi k_1|\sqrt{q^2+M^2}-|y+2\pi k_2|\sqrt{(q-p)^2+M^2}}.
\end{align}
We throw away terms proportional to $p_\mu, p_\nu, p_\lambda$ or $p_\rho$, and get:
\begin{align}
&2\pi\sum_{k_1+k_2=k}\int_0^{2\pi}\d y\int\frac{d^4q}{(2\pi)^4}\Bigg[
\frac{(M^2 + q(q-p) )q_{(\nu}g_{\mu)(\lambda}q_{\rho)}}{\sqrt{q^2+M^2}\sqrt{(q-p)^2+M^2}}\cr
&+q_{(\nu}g_{\mu)(\lambda}q_{\rho)} {\rm Sign}(y-2\pi k_1){\rm Sign}(y+2\pi k_2)\Bigg]e^{-|y-2\pi k_1|\sqrt{q^2+M^2}-|y+2\pi k_2|\sqrt{(q-p)^2+M^2}}.
\end{align}
Computing the sums and integrating over $y$ using the formulas (\ref{AB}) and (\ref{ABSign}), we find:
\begin{align}
&2\pi\int\frac{d^4q}{(2\pi)^4}\Bigg[
\frac{(M^2 + q(q-p) )q_{(\nu}g_{\mu)(\lambda}q_{\rho)}}{\sqrt{q^2+M^2}\sqrt{(q-p)^2+M^2}}\times\cr
&\left(\frac{e^{-2\pi |k| \sqrt{(q-p)^2+M^2}}-e^{-2\pi |k|\sqrt{q^2+M^2}}}{\sqrt{q^2+M^2}-\sqrt{(q-p)^2+M^2}} + \frac{e^{-2\pi |k|\sqrt{q^2+M^2}}+e^{-2\pi |k|\sqrt{(q-p)^2+M^2}}}{\sqrt{q^2+M^2}+\sqrt{(q-p)^2+M^2}} \right)\cr
&+q_{(\nu}g_{\mu)(\lambda}q_{\rho)} \left(-\frac{e^{-2\pi |k| \sqrt{(q-p)^2+M^2}}-e^{-2\pi |k|\sqrt{q^2+M^2}}}{\sqrt{q^2+M^2}-\sqrt{(q-p)^2+M^2}} + \frac{e^{-2\pi |k|\sqrt{q^2+M^2}}+e^{-2\pi |k|\sqrt{(q-p)^2+M^2}}}{\sqrt{q^2+M^2}+\sqrt{(q-p)^2+M^2}}\right)\Bigg].\cr
\end{align}
Now we have to Taylor expand this to get an $O(p^4)$-order contribution. We then integrate over $\d^4q$ at that order. We have to do the same tricks with Wick rotation and replacing products of $q_\mu$ by symmetric combinations of $\eta_{\mu\nu}$:
\begin{align}
&\int \frac{\d^4q}{(2\pi)^4}\left[f_1(q^2)q_\mu q_\rho (p^2)^2 + f_2(q^2)q_\mu q_\rho p^2 (qp)^2 + f_3(q^2)q_\mu q_\rho (qp)^4\right] \to\cr
&-i\int \frac{\d^4q_E}{(2\pi)^4}\left[f_1(q_E^2)\frac{q^2}{4}\eta_{\mu \rho} (p^2)^2 + f_2(q_E^2)\frac{q^4}{24}\eta_{\mu\rho} (p^2)^2 + f_3(q_E^2)\frac{q^6}{64}\eta_{\mu\rho} (p^2)^2\right] +\dots
\end{align}
where the ellipsis represents terms that vanish upon contractions with $h^{\mu\nu}h^{\lambda\rho}$.

So, after Taylor expansion, we get:
\begin{align}
\label{hugeint}
-i\frac{(p^2)^2}{4\pi}\int_0^\infty q^3\d q\Big[
\frac{3 e^{-2 |k| \pi  \sqrt{M^2+q^2}} M^4 q^2}{16 \left(M^2+q^2\right)^{9/2}}+\frac{e^{-2 |k| \pi  \sqrt{M^2+q^2}} k^2 M^6 \pi ^2 q^2}{4 \left(M^2+q^2\right)^{9/2}}+\frac{e^{-2 |k| \pi  \sqrt{M^2+q^2}} M^2 q^4}{6 \left(M^2+q^2\right)^{9/2}}\cr
+\frac{5 e^{-2 |k| \pi  \sqrt{M^2+q^2}} k^2 M^4 \pi ^2 q^4}{12 \left(M^2+q^2\right)^{9/2}}+\frac{73 e^{-2 |k| \pi  \sqrt{M^2+q^2}} q^6}{1536 \left(M^2+q^2\right)^{9/2}}+\frac{77 e^{-2 |k| \pi  \sqrt{M^2+q^2}} k^2 M^2 \pi ^2 q^6}{384 \left(M^2+q^2\right)^{9/2}}\cr
+\frac{e^{-2 |k| \pi  \sqrt{M^2+q^2}} k^4 M^4 \pi ^4 q^6}{96 \left(M^2+q^2\right)^{9/2}}+\frac{13 e^{-2 |k| \pi  \sqrt{M^2+q^2}} k^2 \pi ^2 q^8}{384 \left(M^2+q^2\right)^{9/2}}+\frac{e^{-2 |k| \pi  \sqrt{M^2+q^2}} k^4 M^2 \pi ^4 q^8}{48 \left(M^2+q^2\right)^{9/2}}\cr
+\frac{e^{-2 |k| \pi  \sqrt{M^2+q^2}} k^4 \pi ^4 q^{10}}{96 \left(M^2+q^2\right)^{9/2}}+\frac{3 e^{-2 |k| \pi  \sqrt{M^2+q^2}} |k| M^4 \pi  q^2}{8 \left(M^2+q^2\right)^4}+\frac{e^{-2 |k| \pi  \sqrt{M^2+q^2}} |k| M^2 \pi  q^4}{3 \left(M^2+q^2\right)^4}\cr
-\frac{e^{-2 |k| \pi  \sqrt{M^2+q^2}} |k|^3 M^4 \pi ^3 q^4}{9 \left(M^2+q^2\right)^4}+\frac{73 e^{-2 |k| \pi  \sqrt{M^2+q^2}} |k| \pi  q^6}{768 \left(M^2+q^2\right)^4}-\frac{49 e^{-2 |k| \pi  \sqrt{M^2+q^2}} |k|^3 M^2 \pi ^3 q^6}{288 \left(M^2+q^2\right)^4}\cr -\frac{17 e^{-2 |k| \pi  \sqrt{M^2+q^2}} |k|^3 \pi ^3 q^8}{288 \left(M^2+q^2\right)^4}
\Big].
\end{align}
Doing the same change of variables $x=\sqrt{M^2 + q^2}$ as before and integrating, we get:
\begin{equation}
-i\frac{(p^2)^2}{4\pi}\frac{e^{-2\pi |k| M}}{24\pi |k|}.
\end{equation}
Summing over $k\ne 0$ and using (\ref{tpf}), we obtain: 
\begin{equation}
{\rm Im}\,\F_1^{\rm q} = \frac{1}{16\pi^2}\sum_{k=1}^\infty \frac{e^{-2\pi k M}}{64\times 3 k},
\end{equation}
so, using the fact that $\F_1$ is imaginary at $\alpha^I=0$ and then extending by holomorphy:
\begin{equation}
\F_1^{\rm q} = \frac{1}{16\pi^2} \sum_{k=1}^\infty \frac{i}{64\times 3k} e^{2\pi i k q_I Z^I}.
\end{equation}
This is again compatible with \cite{GVrev}.
\subsubsection*{\emph{A word about} $k=0$}
Just as in the $\F_0$ case, the integral (\ref{hugeint}) is convergent only for $k\ne 0$. The $k=0$ part is again interpreted as a term in the effective action in 5d. And this term then can or cannot contribute to $\F_1$ by the classical dimensional reduction. In \cite{GVrev}, it was shown that the only possible contribution to $\F_1$ from the classical dimensional reduction is of the form $c_{I,2}Z^I$ with real constants $c_{I,2}$. So the only remaining question one could ask here is whether the $k=0$ part of the one-loop answer could contribute by shifting the values of these $c_{I,2}$. 

The real part of $\F_1\propto c_{I,2}Z^I$ enters the 4d interaction $\int c_{I,2} \alpha^I \Tr( R\wedge R)$, which comes from a Chern-Simons interaction in 5d of the form $\int c_{I,2} V^I\wedge \Tr( R\wedge R)$. The imaginary part of $\F_1$ corresponds to the 4d interaction $\int \sqrt{g}\d^4x\,c_{I,2}h^I R^2$, which apparently lifts to the 5d interaction of the form $\propto\int\sqrt{G}\d^5x\, c_{I,2}h^I R^2$. While the meaning of the latter term is not entirely clear, the 5d Chern-Simons term actually looks familiar. As explained in \cite{GVrev}, it can be lifted even further, to the 11d action. Its 11d origin is an interaction $\frac{1}{(2\pi)^4}\int C\wedge \left[ \frac{1}{768}(\Tr R^2)^2 - \frac{1}{192}\Tr R^4\right]$ (where the powers of $R$ are with respect to the wedge product). This interaction was discovered in \cite{origin} due to its role in the anomaly cancelation in M-theory. This suggests that $c_{I,2}$ cannot be shifted -- one can run the same anomaly argument in 5d to determine the values of $c_{I,2}$. Another evidence that quantum corrections cannot shift $c_{I,2}$ appears if we turn on holonomies $\alpha^I$. We know that they appear in a diagram computation only through the factors $e^{2\pi i k q_I\alpha^I}$, which means that the term $\int c_{I,2}\alpha^I\Tr(R\wedge R)$ (which has to be generated at $\alpha^I\ne 0$ background) cannot be shifted. Thus $c_{I,2}$ is not actually shifted by the $k=0$ part of the one-loop answer, and it is enough to consider only $k\ne 0$ terms.

\section{Discussion}\label{disc}
We have computed the contribution of a single light hypermultiplet to $\F_0$ and $\F_1$. As was explained in \cite{GVrev} and originally noticed in \cite{GV1}, to get a contribution from all of the massless multiplets in the theory (that is, hypermultiplets, vector multiplets and the gravity multiplet), one just has to multiply the contribution of a massless hypermultiplet by $-\chi(Y)/2$, where $\chi(Y)$ is Euler characteristic of the Calabi-Yau $Y$. The massless hypermultiplet contribution is a massless limit of what we have computed here.

Note that the superparticle description, which was advocated in Section 3 of \cite{GVrev} (and which is a perfect choice for massive BPS multiplets), does not have a sensible massless limit, even though in the answer one can formally take mass to zero. That is why the field theoretic description was essential for the complete picture. For $\g\ge 2$, the field-theoretic computation of $\F_\sg$ was described in Section 4 of \cite{GVrev}. The field-theoretic computation of $\F_0$ and $\F_1$ is presented in this paper, thereby completing the physical treatment of the GV formula. There are several other points we want to make.

One point is about possible improvements of our computation. One could try to generalize the field theoretic derivation by finding an alternative and probably more natural one. Notice that in \cite{GVrev}, for $\g\ge 2$, one had to perform a Poisson resummation to bring field-theoretic answer into a useful form, when it is presented as a sum over the winding number $k$. But in the derivation we have for $\F_0$ and $\F_1$, we got the answer as a sum over $k$ without any Poisson resummation. Thus one could ask if it is possible to generalize the approach we have taken here to $\g\ge 2$.

It is quite clear how to generalize our computation of $\F_1$. Since the terms $\F_\sg$ give rise to interactions of the form $\F_\sg(X)(R^-)^2({\sf W}^{-\,2})^{\sg-1} + c.c.$, one again could try to determine $\F_\sg$ by computing the two point function of stress-energy tensor, but now in a flat background with the graviphoton field ${\sf W}^-$ (or ${\sf T}^-$ in the 5d language) turned on. While such an approach is completely feasible, it seems to be much harder computationally than the approach of Section 4 of \cite{GVrev}. However, it may well turn out that the actual computation will be easier than we expect.

Another possible approach is to generalize the computation of $\F_0$. In this case we notice that turning on a constant graviphoton background ${\sf T}^-$ (i.e. considering the supersymmetric G\"odel solution of \cite{sol5d}, as explained in \cite{GVrev}) effectively deforms the 4d prepotential by the 4d graviphoton ${\sf W}^-$:
\begin{equation}
\tilde{\F}_0(X) = \sum_{\sg=0}^\infty \F_\sg(X) ({\sf W}^{-\, 2})^\sg,
\end{equation}
so that if we treat ${\sf W}^-$ not as a field but rather as a parameter in the action, the kinetic term for scalars will receive ${\sf W}^-$-dependent deformations. One can reconstruct $\F_\sg$ from the knowledge of this deformed kinetic term. And to compute the deformed kinetic term, all we need to do is to compute the two-point function of mass terms (as in the $\F_0$ computation in Section \ref{F0} of this paper), but in the background with the constant graviphoton field turned on. Again, it seems presently that such a computation will be much more complicated then what we have so far. But if it turns out to be simple, then it will be a nice approach to compute $\F_\sg$ for all $\g\ge 0$ at once.

Finally, we note that the results of the one-loop computation are actually exact. This one-loop exactness follows in the usual way from holomorphy. If we go beyond quadratic order in the action and thus consider higher-loop corrections, they will be multiplied by extra powers of the mass $M=q_I h^I$, which will not be balanced by extra powers of holonomies $q_I\alpha^I$, and thus will violate holomorphy.

\section*{Acknowledgments}
I would like to thank E.Witten for valuable discussions and help. I also thank V.Mikhaylov for valuable discussions.

\end{document}